\documentclass[reprint, amsmath,amssymb, floatfix, pre, showpacs]{revtex4-1}
\usepackage{graphicx}
\usepackage{amssymb}
\usepackage{amsmath}
\usepackage{color}
\usepackage{psfrag}
\usepackage{epsfig}
\usepackage{bbm}
\usepackage{bm}
\usepackage{dsfont}
\usepackage{csquotes}

\newcommand{\im}{\mathrm{i}}

\newcommand{\rb}[1]{\left( #1 \right)}
\newcommand{\mb}[1]{\mathbf{#1}}

\newcommand{\beq}{\begin{eqnarray}}
\newcommand{\eeq}{\end{eqnarray}}

\newcommand{\eq}[1]{Eq.~(\ref{#1})}

\begin{document}
\title{Shear-induced laning transition in a confined colloidal film}
\author{Sascha~Gerloff, Tarlan~A.~Vezirov and Sabine H.~L.~Klapp}
\affiliation{
  Institut f\"ur Theoretische Physik, Hardenbergstr. 36, Technische Universit\"at Berlin, D-10623 Berlin, Germany
}

\date{\today}
\begin{abstract}
Using Brownian dynamics (BD) simulations we investigate a dense system of charged colloids exposed to shear flow in a confined (slit-pore) geometry.
The equilibrium system at zero flow consists of three, well-pronounced layers with square-like crystalline in-plane structure.
We demonstrate that, for sufficiently large shear rates, the middle layer separates into two sublayers where the particles organize into moving lanes with opposite velocities.
The formation of this micro-laned state results in a destruction of the applied shear profile.
It has a strong impact not only on the structure of the system, but also on its rheology as measured by the stress tensor.
At higher shear rates we observe a disordered state and finally a recrystallization reminiscent of the behavior of bilayer films.
We expect the shear-induced laning to be a generic feature of \emph{thin} films with three or more layers.
\end{abstract}
\pacs{{\tt 82.70.Dd, 83.50.Ax, 05.70.Ln, 82.70.Kj}}
\maketitle
\section{Introduction \label{SEC:INTRO}}
Dense colloidal suspensions subject to strong spatial confinement can form solid-like structures not seen in their bulk counterparts.
A well-established example are hard-sphere-like colloidal particles between two parallel plates with a distance $L_{\mathrm{z}}$ of the order of the particles diameter $d$ \cite{Fortini2006, Oguz2012, Curk2012,Khadilkar2016}.
Depending on the commensurability of $L_{\mathrm{z}}$ and $d$ (as well as the chemical potential) one observes the formation of $n$ layers with either square or hexagonal lattice symmetry within the plane parallel to the plates (with alternating order $n\square$, $n\triangle$, $(n+1)\square$, $\ldots$) \cite{Fortini2006, Oguz2012}.
Moreover, at high densities exotic structures such as buckled and rhombic crystals do occur \cite{Fortini2006}.
While the equilibrium behavior of strongly confined suspensions is well understood for many colloidal and molecular systems, the interplay between particle packing and shear flow remains surprisingly elusive.
This contrasts the fact that sheared films are of major relevance in many industrial processes involving surface coatings, lubricants and microfluidic devices.

A number of recent experimental and numerical studies has focused specifically on \emph{hydrodynamic} effects occurring due to the interplay of confinement and flow \cite{Aerov2015, Cohen2004, Mehrabadi2016, Mackay2014, Myung2013, Wilms2012, Isa2009}.
Hydrodynamic effects have also been considered in the context of slit-pores with modulated pore widths \cite{Genovese2011, Martens2014, Marconi2012, Zimmermann2016} and confined active particles \cite{Apaza2016, Kreuter2013, Wioland2016}.
However, these studies typically consider dilute systems and relatively wide gaps involving a bulk-like region (at zero flow) in the middle region \cite{Mackay2014, Myung2013, Wilms2012}.
Another set of studies considers the flow of non-Brownian particles (no thermal fluctuations), including their frictional properties as function of $L_{\mathrm{z}}$ \cite{Fornari2016, Yeo2010}.

In the present paper we are interested in the shear-induced behavior of a thin film of Brownian spheres which, already in equilibrium, forms a crystalline lateral structure with square symmetry.
Specifically, we consider the case of $n=3$ layers.
The system is studied on the basis of overdamped Brownian dynamics (BD) simulation without hydrodynamic interactions (following various earlier studies on dense systems under flow \cite{Besseling2012, Cerda2008, Lander2013, Messina2006}).
The particles interact via screened Coloumb interactions (matched to a real, silica system \cite{Grandner2008, Klapp2008}), and the confining walls are considered as smooth on the particle scale.
Shear is then imposed via a force acting on each particle, which only depends on the distance relative to the walls.

The present investigation extends earlier numerical and theoretical studies by some of us where we focused on shear-induced transitions in bilayer systems \cite{Vezirov2013, Vezirov2015} and, in particular, their description within the Frenkel-Kontorova model for solid friction \cite{Gerloff2016}.
From the perspective of a shear-driven bilayer, the trilayer film can be seen as a first step towards the third dimension.
Moreover, as we will demonstrate, the trilayer films displays a novel effect which was already briefly mentioned in Refs.~\citenum{Vezirov2015,Gerloff2016} and which we think is typical for sheared \emph{thin} films with three or more layers: This effect consists of the formation of a state where the particles in the middle layer organize into \emph{lanes} along the flow direction.
Laning is a typical non-equilibrium effect which occurs quite generically in binary systems of particles driven in opposite directions (such as in dusty plasmas \cite{Morfill2006, Suetterlin2009}, granular matter \cite{Ciamarra2005}, in pedestrian and ant dynamics \cite{Helbing1995, Helbing2005, Couzin2003} as well as in oppositely charged colloids driven by electric field \cite{Comiskey1998, Gelinck2004}).
In the present case, the driving force is represented by the shear flow generating competing effects on the particles in the middle layer.
Moreover, the laning is accompanied by a transformation of the (originally flat) middle layer into two sublayers.
This is in sharp contrast to the behavior of bilayers (and also of bulk suspensions) where the layers always remain flat \cite{Besseling2012}.
In our system, the laned state occurs at shear rates in between the initial square state and the subsequent shear melting.
Interestingly, a similar effect (stripe formation and related "buckling") has been reported in the context of experiments \cite{Cohen2004} of strongly confined hard sphere suspensions.

The paper is organized as follows.
In the subsequent section~\ref{SEC:MODS}, we describe our model system as well as some details of the computer simulations.
Numerical results are presented in Sec.~\ref{SEC:RES}, where we start by describing the overall dynamical behavior, followed by a detailed discussion of the shear-induced laning transition, the resulting single-particle dynamics, and the implications for the shear stress.
Finally we summarize and conclude in Sec.~\ref{SEC:CONC}.

\section{Model and simulation details}
\label{SEC:MODS}
Following previous studies \cite{Vezirov2013, Vezirov2015} we consider a model colloidal suspension where two spherical particles (diameter $d$) with distance $r_{ij}$ interact via a combination of a repulsive Yukawa potential,  $u_{\mathrm{Yuk}}(r_{ij}) = W\exp\left[-\kappa r_{ij}\right]/r_{ij}$ and a repulsive soft-sphere potential $u_{\mathrm{SS}}(r_{ij})=4 \epsilon( d/r_{ij})^{12}$.

The interaction parameters are set to $W/(k_{\mathrm{B}}T d)\approx 123$ (where $k_{\mathrm{B}}T$ is the thermal energy) and the inverse Debye screening length is set to $\kappa\, d \approx 3.2$ (for details, see Ref. \cite{Grandner2008, Klapp2008}).
Spatial confinement is modeled by two plane parallel, smooth, uncharged surfaces separated by a distance $L_{\mathrm{z}}$ along the $z$ direction and of infinite extent in the $xy-$ plane. We employ a purely repulsive fluid-wall potential \cite{Vezirov2013}
\begin{equation}\label{EQ:fluid-wall potential}
  u_\text{wall}^{\rb{\pm} } \rb{z_i} = \frac{4\pi \epsilon_w}{5} \rb{ \frac{d}{ z_i \pm L_z/2 } }^9\text{,}
\end{equation}
where $z_i$ is the $z$-position of particle $i$.
The dimensionless coupling parameter $\epsilon_w/k_BT$ is set to one.
This choice of potential is motivated by previous investigations of the equilibrium structure, where we found a very good agreement with experiments \cite{Grandner2008, Klapp2008}.
Our investigations are based on standard BD simulations in three dimensions,  where the position of particle $i$ is advanced according to \cite{Ermak1975}
\begin{equation}
\mathbf{r}_{i}(t+\delta t) = \mathbf{r}_{i}(t)+\frac{D_{0}}{k_{B} T}\mathbf{F}_{i}(t)\delta t+\delta \mathbf{W}_{i}+\dot{\gamma}z_{i}(t)\delta t \mathbf{e}_{x}.
\label{EQ:Eqmot}
\end{equation}
Here, $\mathbf{F}_{i}$ is the total conservative force acting on particle $i$ and $\delta \mathbf{W}_{i}$ denotes a random Gaussian displacement with zero mean and variance of $2D_{0}\delta t$ for each Cartesian component.
The friction constant is $(D_{0}/k_{B} T)^{-1}$, where $D_{0}$ is the short-time diffusion coefficient.
The timescale of the system is set to $\tau = d^{2}/D_{0}$, which defines the so-called Brownian time.
We impose a linear shear profile [see last term in Eq.~(\ref{EQ:Eqmot})] representing flow in $x$- and gradient in $z$-direction, characterized by uniform shear rate $\dot{\gamma}$ \cite{Messina2006, Rottler2007}.
We note, however, that despite the application of a linear shear profile, the real, steady-state flow profile can be non-linear \cite{Delhommelle2003}.
Furthermore, to reduce the computation time we neglect hydrodynamic interactions.
We focus on strongly confined ($L_{\mathrm{z}}=3.2 d$) and dense ($\rho  d^3=0.85$) system with $1587$ particles.
The resulting lateral width of the simulation cell then follows as $L\approx 24.2 d$.
Periodic boundary conditions were applied in flow ($x$) and vorticity ($y$) direction.
The time step is set to $\delta t=10^{-5} \tau$.
After the equilibration period of $10^{7}$ steps (i.e. 100$ \tau$~) the system relaxes (in agreement with our expectations \cite{Klapp2008, Vezirov2015}) in a trilayer with square in-plane order.
Then the shear force is switched on.
After starting the shearing, the simulation was carried on for an additional period of 100$ \tau$.
During this time the steady state is reached.
Only after these preparations we started to analyze the considered systems.
%
\section{Numerical results \label{SEC:RES}}
In this section we present our numerical results for the dynamical behavior of the shear-driven trilayer.
We start by reviewing the overall behavior, followed by a detailed discussion of the novel, \emph{micro-laned} state of the middle layer, the associated single-particle dynamics and the resulting rheological behavior.
\subsection{Overview \label{SEC:TO}}
\begin{figure}
\includegraphics[width=1.0\linewidth]{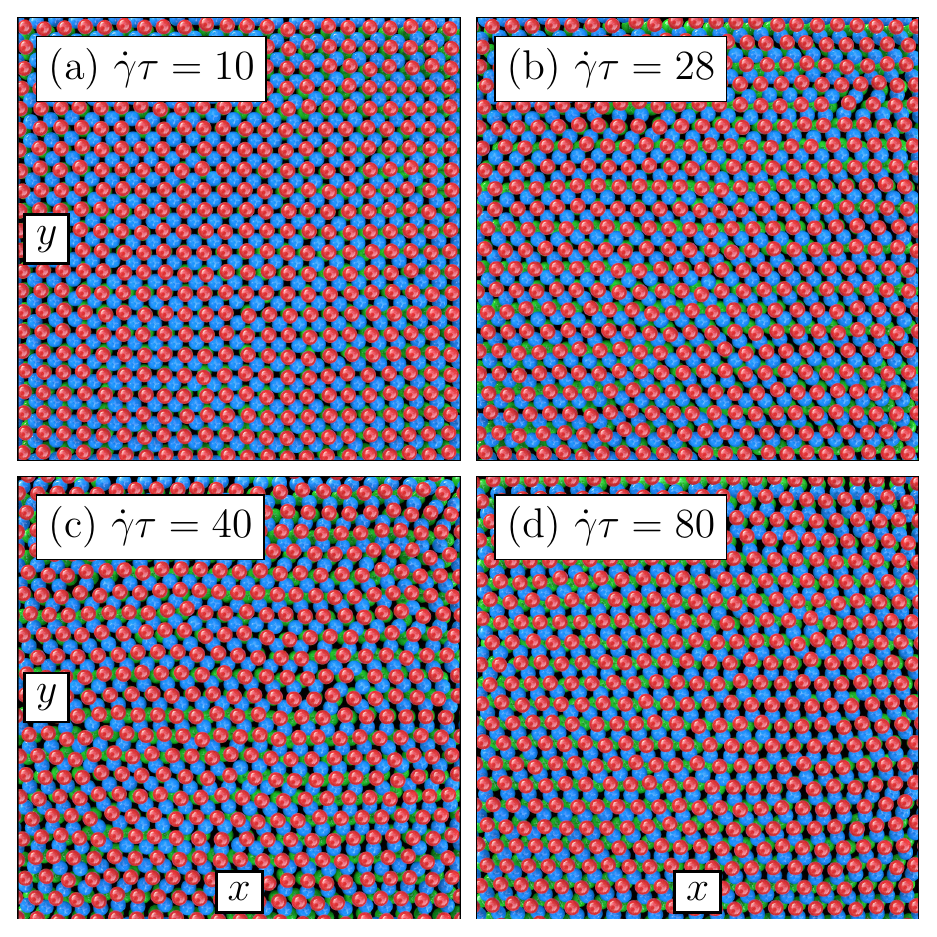}
\caption{(Color online) (a)-(d) Simulation snapshots of the $xy$-plane at $\rho d^3=0.85$ and $L_z=3.2 d$ for different shear rates. The red (blue, green) circles represent particles of the upper (middle, lower) layer.}
\label{FIG:Snapshot1}
\end{figure}

As a first overview of the shear-induced structural changes we present in Fig.~\ref{FIG:Snapshot1} BD simulation snapshots at representative values of the dimensionless shear rate $\dot\gamma\tau$.
We can identify four different states.
The first one is a laterally crystalline state with \emph{square} in-plane ordering in each of the three layers.
This state corresponds to the equilibrium configuration ($\dot\gamma=0$) at the thermodynamic conditions considered, and it persists at small shear rates such as $\dot\gamma\tau=10$ [see Fig.~\ref{FIG:Snapshot1}(a)].
The degree of (local) quadratic translational order in each layer is quantified by the order parameter $\Psi_{n}$ with $n=4$, with $\Psi_n$ being generally defined as
\begin{align}
\Psi_{n}=\left\langle \frac{1}{N_{L}}\sum_{i=1}^{N_{L}} \frac{1}{N_{i}^{b}} \bigg| \sum_{j=1}^{N_{i}^{b}}\exp(\im n\theta_{j}) \bigg| \right\rangle.
\label{EQ:Order}
\end{align}
In \eq{EQ:Order}, $N_{L}$ describes the instantaneous number of particles and $N_{i}^{b}$ the number of neighbors of particle $i$ in the considered layer (the position of each layer can be extracted from the density profile, see below).
The value of $N_{i}^{b}$ was calculated from the intralayer pair correlation function by determining the radius related to the first minimum of the intralayer pair correlation function \cite{Vezirov2013}.
Perfect square or hexagonal order is characterized by $\Psi_{4/6}=1$, respectively.

\begin{figure}
\includegraphics[width=1.0\linewidth]{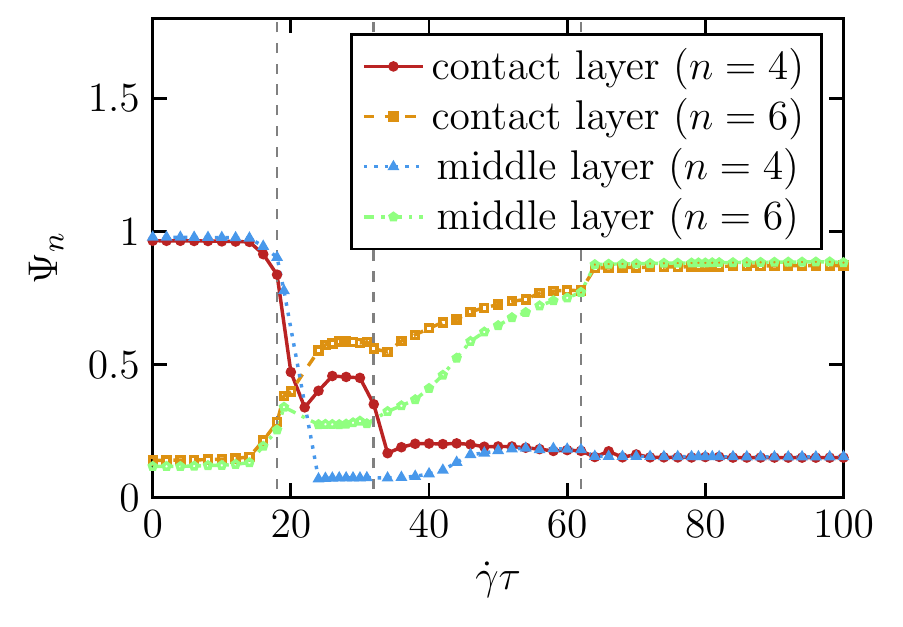}
\caption{(Color online) In-plane order parameter for square ($\Psi_4$) and hexagonal ($\Psi_6$) symmetry as a function of the dimensionless shear rate $\dot \gamma \tau$. The order parameter was calculated separately for the middle and the contact layers.}
\label{FIG:Order}
\end{figure}

\begin{figure}
\includegraphics[width=1.0\linewidth]{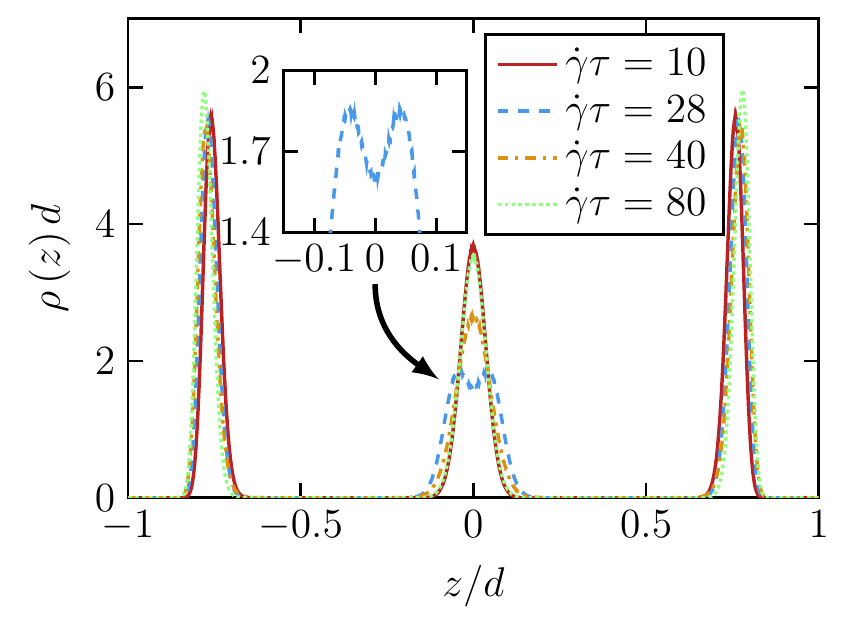}
\caption{(Color online) Density profiles along the shear gradient (and confinement) direction for different shear rates. In the inset, the middle peak of the density profile for $\dot\gamma\tau = 28$ is magnified, revealing the double peak structure.}
\label{FIG:Density}
\end{figure}

Results for $\Psi_4$ and the hexagonal order parameter $\Psi_6$ for the individual layers as functions of the shear rate are plotted in Fig.~\ref{FIG:Order}.
For completeness, we also present in Fig.~\ref{FIG:Density} the density profile in $z$-direction, $\rho(z)=\langle N(z)/N\Delta z\rangle$ (where $N(z)$ is the instantaneous number of particles at a given distance $z$ and $\Delta z$ defines the bin width).
In the initial state, all layers are characterized by a high degree of square-like order (and, correspondingly, very small values of $\Psi_6$).
This changes at $\dot\gamma\tau\approx 18$, where we observe a sudden decrease of $\Psi_4$.
A visualization of the system's structure in the subsequent state is given in Fig.~\ref{FIG:Snapshot1}(b) suggesting, at first sight, a substantial loss of in-plane order as compared to the initial state.
However, as we will discuss in more detail in Sec.~\ref{SEC:SL}, this state is characterized by a "splitting" of the middle layer into two sublayers, where the particles in the two sublayers organize into lanes.
Therefore, the top view provided in Fig.~\ref{FIG:Snapshot1}(b) provides only partial information.
The splitting can already be seen from the density profile (see Fig.~\ref{FIG:Density}) which displays, at $\dot\gamma\tau=28$, a double peak in the middle of the slit.
This contrasts the single middle peak observed at other shear rates.
The third state [see~Fig.~\ref{FIG:Snapshot1}(c)] has an overall disordered appearance, with the degree of hexagonal order being somewhat larger than the values of $\Psi_4$.
The corresponding density distribution (see Fig.~\ref{FIG:Density}) is characterized by an unsplitted, yet rather broad peak at the middle position.
This "molten" state resembles the one seen in shear-driven \emph{bilayer} systems at shear rates directly beyond the quadratic regime \cite{Vezirov2013}.
This correspondence also holds for even higher shear rates where the trilayer system (in analogy to the bilayer) recrystallizes into a state with hexagonal in-plane order in all layers (see~Fig.~\ref{FIG:Snapshot1}(d) and Fig.~\ref{FIG:Order}).

We can thus conclude that while the limiting behavior of the trilayer at low and high shear rates, respectively, is identical to that in bilayer systems \cite{Vezirov2013}, the breakdown of ordering at
intermediate shear rates shows pronounced differences.
In the next subsection we analyze in more detail the structure in this "micro-laned" state.
\subsection{Laning transition\label{SEC:SL}}
\begin{figure}
\includegraphics[width=1.0\linewidth]{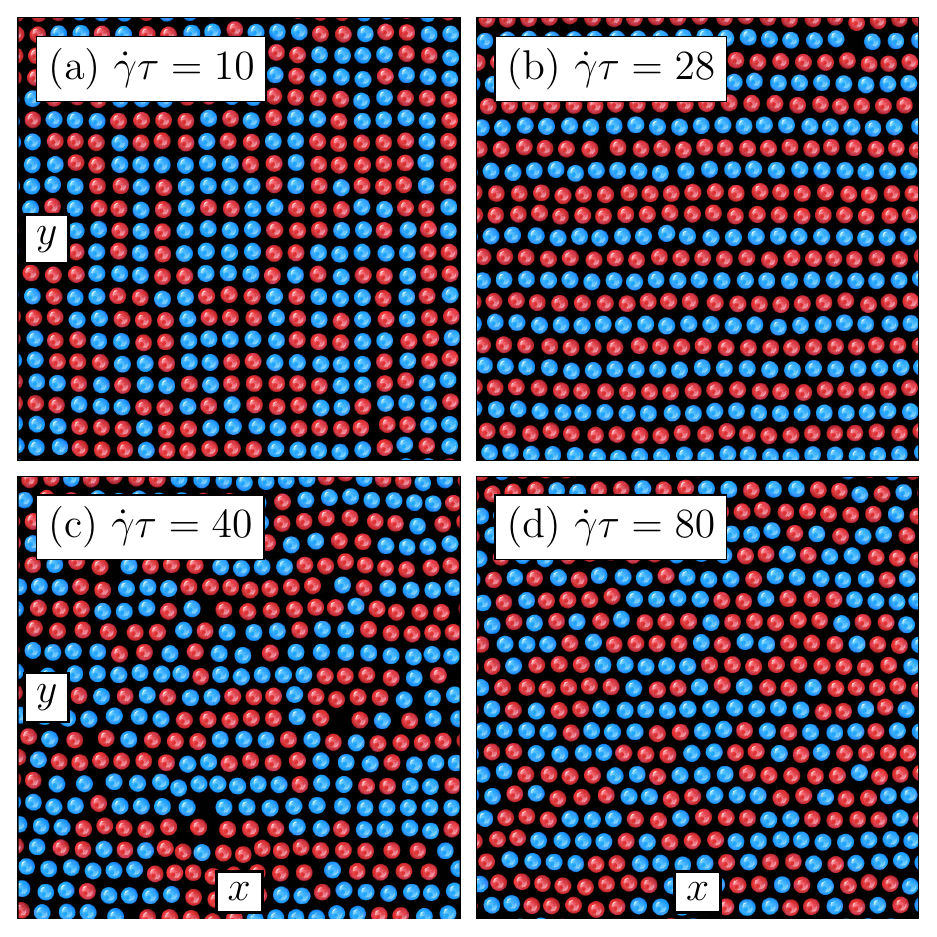}
\caption{(Color online) (a)-(d) BD simulation snapshots of the middle layer at $\rho d^3= 0.85$ and $L_z = 3.2 d$ for different shear
rates. The blue (red) circles represent particles with $z>0$ ($z<0$).}
\label{FIG:Snapshot2}
\end{figure}

In order to better understand the nature of laned state we present in Fig.~\ref{FIG:Snapshot2} snapshots only of the middle layer at the same shear rates considered already in Fig.~\ref{FIG:Snapshot1}.
We note that, independent of the precise value of $\dot\gamma \tau$, the middle layer has a certain width involving particles located at $z$-values slightly above or below the middle plane at $z=0$.
In Fig.~\ref{FIG:Snapshot2} we distinguish between these particle by different colors.
The snapshots at low and high shear rates reveal, as expected, the occurrence of quadratic [Fig.~\ref{FIG:Snapshot2}(a)] and hexagonal [Fig.~\ref{FIG:Snapshot2}(d)] order, respectively, with the colors of the particles suggesting an essentially random distribution of positive and negative $z$-values within the middle plane.
This mixed distribution also holds in the shear-molten state [Fig.~\ref{FIG:Snapshot2}(c)].
In contrast, Fig.~\ref{FIG:Snapshot2}(b) clearly demonstrates a "micro" phase-separation of the particles: The two different types arrange into narrow "lanes" along the flow ($x$-) direction, with the quadratic order between particle of different lanes being much less pronounced than in the initial quasi-equilibrium state.
This separation explains the splitting of the density peak observed at $\dot\gamma\tau=28$ (see Fig.~\ref{FIG:Density}), as well as the accompanying decrease of the order parameter $\Psi_4$ (Fig.~\ref{FIG:Order}).

From a physical point of view the occurrence of the lanes is maybe not surprising: We recall that, in the present shear geometry, particles at equal distance from the middle plane experience an \emph{oppositely} directed shear force via the externally controlled flow profile.
The occurence of a "laning transition" in binary systems of oppositely driven particles is indeed a well-established phenomenon which has been observed in a wide variety of systems including colloids and pedestrians~\cite{Dzubiella2002, Helbing2005}.

Inspired by the numerous studies of laning in colloidal systems we here define a laning order parameter (see, e.g.,~\cite{Dzubiella2002}) addressing the particles in the middle layer.
We first assign to each particle the parameter $\phi_{i}$, which is chosen to be $1$ if the lateral distance $|y_{i}-y_{j}|$ to all particles of the adjacent \emph{sublayer} $j$ is larger than the average distance $r_{lane}=\rho^{-1/3} / 2$.
Otherwise, $\phi_{i}$ is set to $0$. The laning order parameter is then defined as

\begin{align}
\Phi=\left\langle \frac{1}{N_{L}^\text{mid}} \sum_{i=1}^{N_{L}^\text{mid}}\phi_{i}\right\rangle.
\label{laning}
\end{align}

In Eq.~\eqref{laning} $N_{L}^\text{mid}$ denotes the number of particles within the middle layer.

\begin{figure}
\includegraphics[width=1.0\linewidth]{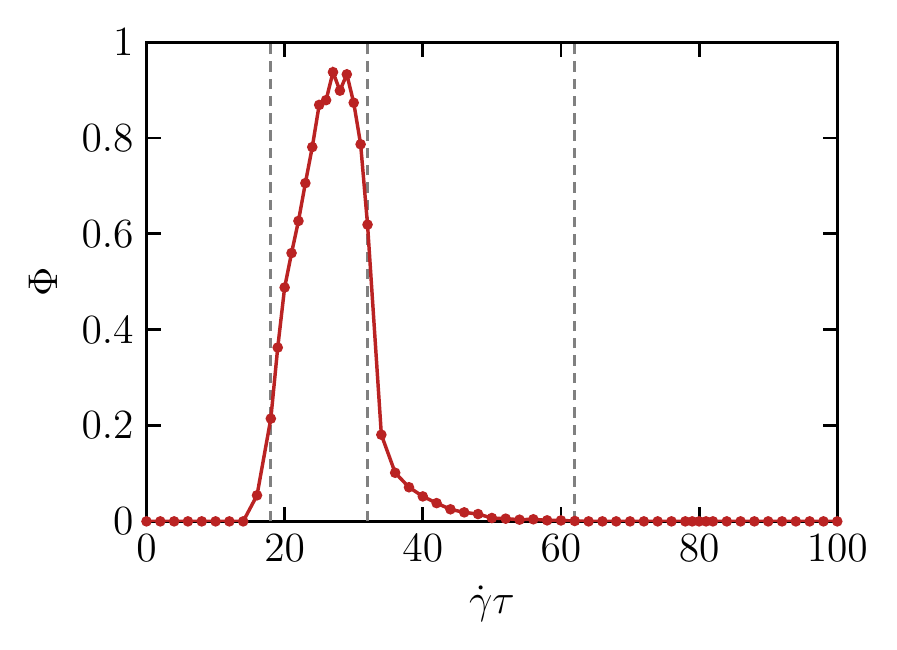}
\caption{(Color online) Laning order parameter within the middle layer as a function of the shear rate.}
\label{FIG:LaneOrder}
\end{figure}

Numerical results for $\Phi$ as function of the shear rate are plotted in Fig.~\ref{FIG:LaneOrder}.
One clearly recognizes a regime of shear rates ($18\lesssim\dot \gamma  \tau\lesssim32$) with strong degree of laning, consistent with the observation in Fig.~\ref{FIG:Snapshot2}(b).
We note that the results for $\Phi$ somewhat depend on the system size, as test simulations reveal.
However, we have checked that the general effect remains even for much larger systems ($N=6627$).

The separation of the system into lanes also explains the relatively low values of the translational order parameters
in this state: The results in Fig.~\ref{FIG:Order} have been obtained by averaging over the entire layers which, for the middle layer,
clearly yields distorted results.
\subsection{Single-particle dynamics \label{SEC:SD}}
\begin{figure}
\includegraphics[width=1.0\linewidth]{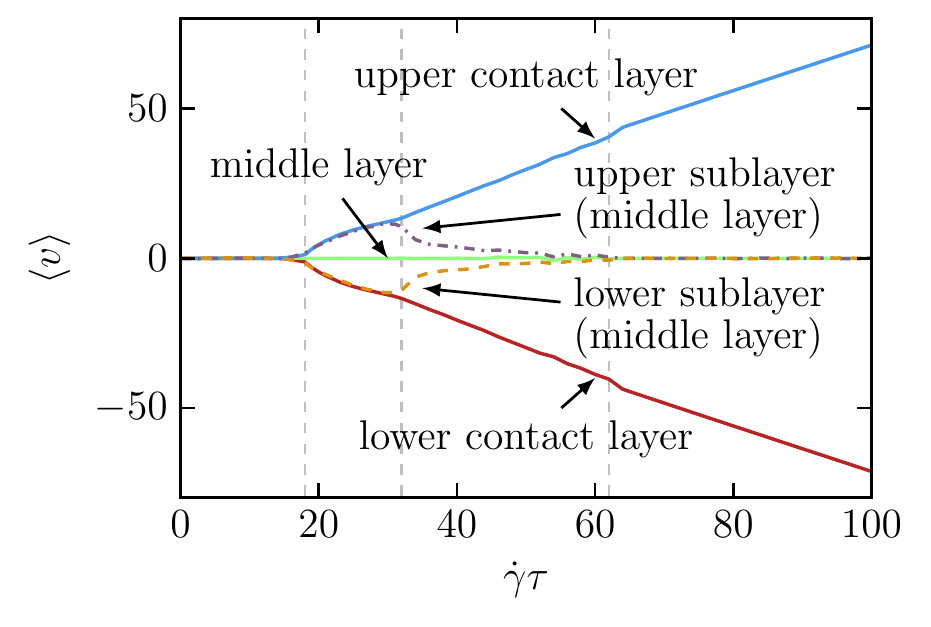}
\caption{(Color online) Velocity of the center of mass of all three layers as well as the sublayers of the middle layer.}
\label{FIG:Velocity}
\end{figure}

We now turn to dynamical aspects associated with the shear-induced structural behavior discussed so far.
To start with, we consider in Fig.~\ref{FIG:Velocity} the velocity of the center of mass of each layer (and sublayer, where appropriate).

The initial, square state is characterized by zero motion of all three layers; in other words, the particles are "locked" in the regular potential valleys formed by the neighboring layers.
On increasing the shear rate towards the laned, molten and hexagonal state the two outer layers start to move, while the \emph{average} velocity in the middle layer remains zero.
This is expected in view of our shear geometry, which implies that the shear force vanishes at $z=0$.
Contrary to this average motion, however, the sublayers formed in the laned state do move, as Fig.~\ref{FIG:Velocity} reveals.
In particular, the corresponding velocities are close to those of the outer layers.
Thus, the velocity profile within the middle layer transforms into two "domains" with opposite velocities.
In the molten state, we still observe sublayer motion but the corresponding velocities "decouple" from those of the outer layers (reflecting a gradual disappearance of lanes).
As a consequence, the sublayer velocities decrease to zero upon approaching the hexagonal state.
We thus see that the laned and molten state, which were hardly distinguishable from their configurations (see Fig.~\ref{FIG:Snapshot1}), differ clearly in their velocity profiles.

\begin{figure}
\includegraphics[width=1.0\linewidth]{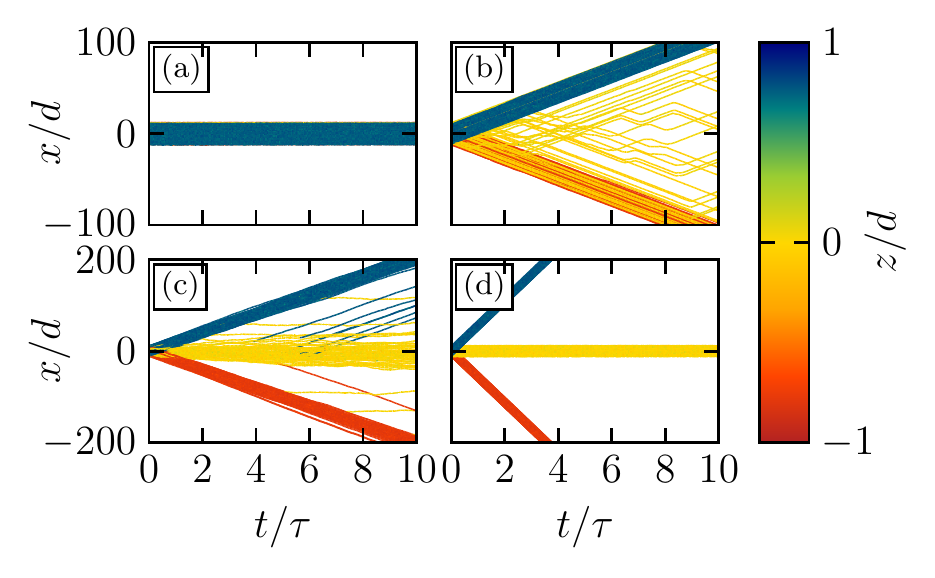}
\caption{(Color online) Trajectories of arbitrarily chosen particles at different shear rates corresponding to the four states.
The colors indicate the position in $z$-direction (see color bar on the right).}
\label{FIG:Trajectory}
\end{figure}

Further information on the dynamics on the particle level is provided by investigating typical trajectories.
Specifically, we consider in Fig.~\ref{FIG:Trajectory} the position of typical particles in flow ($x$-) direction.
The colors indicate the corresponding $z$-position.
The thick horizontal line in part (a), which is composed of aligned trajectories of multiple particles, reflects the "locking" of the particles in their lattice position at low shear rates.
In the other extreme case [hexagonal state, part (d)] we observe uniform flow behavior of the particles in the outer layers, as well as the rest of the ones in the middle plane.
Regarding the intermediate states, the trajectories reflect some degree of non-uniform behavior. In particular, we observe in the lane states trajectories with oscillations around $x=0$ at early times; these oscillations correspond to particles hopping between the sublayers of the middle-plane (and thus, hopping back and forth).
In the molten state the behavior becomes more uniform, in particularly, a large number of particles in the middle plane rests. This indicates that the system approaches a uniform velocity profile similar to the hexagonal state.

\begin{figure}
\includegraphics[width=1.0\linewidth]{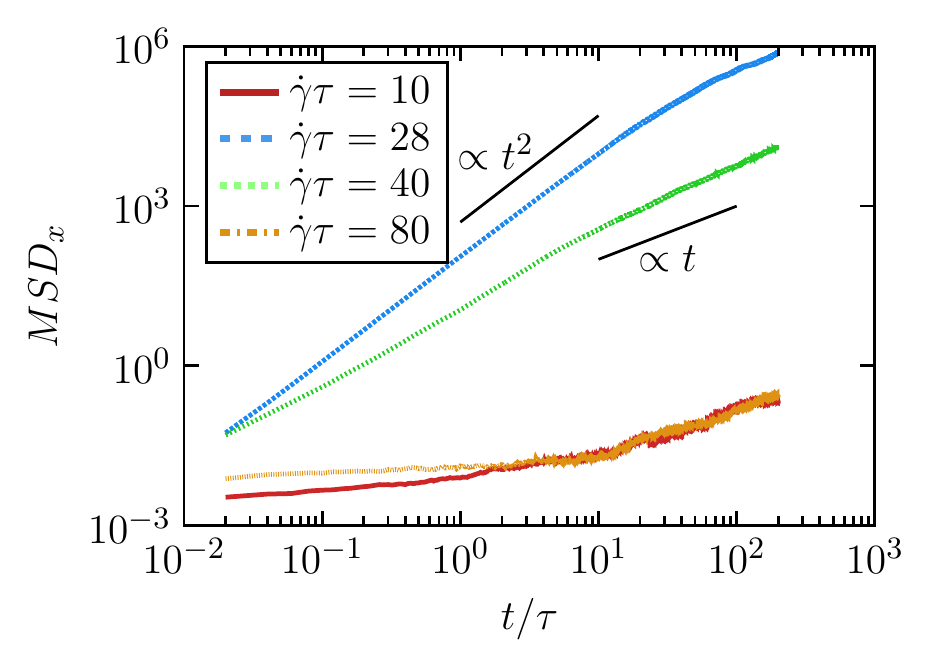}
\caption{(Color online) Mean squared displacement in flow direction relative to the center of mass of the particles in the middle layer at different shear rates.}
\label{FIG:Diffusion}
\end{figure}

A further characteristic dynamical feature of each state is the mean-squared displacement (MSD) relative to the center of mass.
Here we focus on the MSD in the middle layer, since this layer shows the greatest variety in order and dynamics.
Taking into account that the center of mass velocity in this layer is zero, and specializing on the flow ($x$-) direction we define the MSD according to
\begin{align}
MSD_{x}=\left\langle\frac{1}{N_{L}^\text{mid}}\sum_{i=1}^{N_{L}^\text{mid}} \left(x_{i}(t)-x_{i}(0)\right)^{2}\right\rangle,
\label{velsq}
\end{align}
where $N_{L}^\text{mid}$ is the number of particles in the middle layer.
Results for the MSD in the four different states are plotted in Fig.~\ref{FIG:Diffusion}.
In the square ($\dot \gamma \tau=10$) and hexagonal ($\dot \gamma \tau=80$) state, respectively, the MSD quickly develops a plateau, reflecting the persistent trapping of the particles at the sites of the square or hexagonal lattice.
The corresponding long-time limit of the MSD remains smaller than $1d$, consistent with the fact the lattice constants in both cases are approximately of the size of the particle diameter.

\begin{figure}
\includegraphics[width=1.0\linewidth]{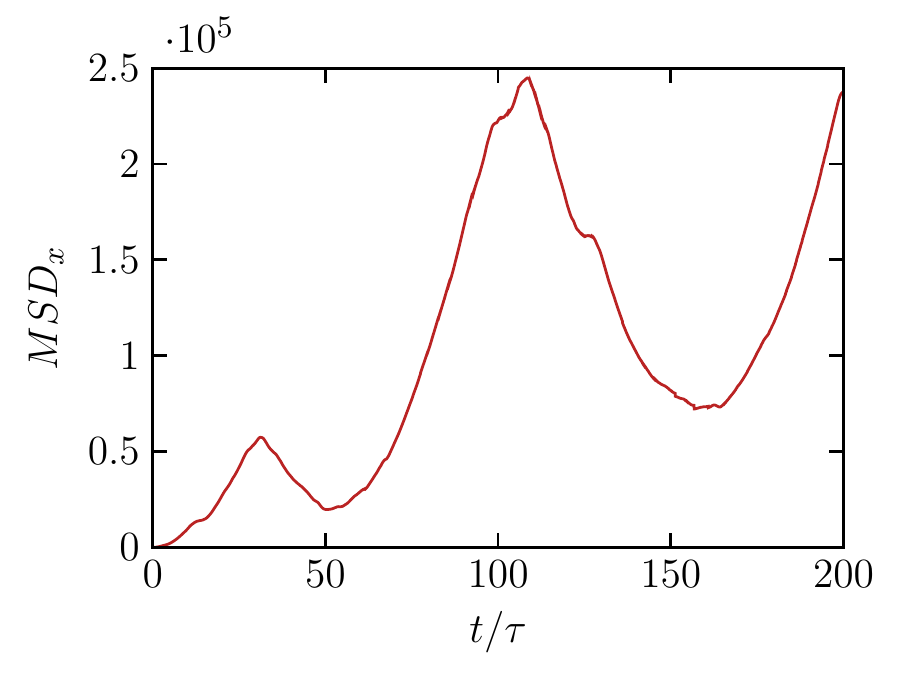}
\caption{(Color online) MSD$_x$ of one single system at $\dot \gamma \tau= 28$. For the calculation only the particles from the middle layer were considered.}
\label{FIG:Diffusion_single}
\end{figure}

In the shear-molten state the long-time behavior of the MSD become diffusive (i.e., linear in time) at long times, as expected in a liquid-like state in the absence of any center-of-mass motion.
The intermediate dynamics can be further analyzed in terms of an effective single-particle model describing a particle in a harmonic potential (mimicking the spatial confinement in $z$-direction) \cite{Vezirov2013}.
Here we rather focus on specific features of the laned state.
The corresponding MSD behaves indeed differently.
In particular, we observe a ballistic regime $\propto t^2$ at intermediate times.
Such a ballistic time-dependence also occurs for a free particle under shear flow~\cite{Orihara2011}, it reflects the presence of net (center-of-mass) motion.
In the present case, the motion is performed by the two sublayers in which the middle layers split (see the non-zero velocities of the sublayers in Fig.~\ref{FIG:Velocity}).
Having this in mind, it seems somewhat surprising that the MSD at later times becomes diffusive again.
Recall, however, that the MSD defined in Eq.~(\ref{velsq}) involves \emph{all} particles in the middle layer.
This includes, in particular, particles jumping between the sublayers.
On the level of the MSD, these jumps (i.e., spontaneous changes of the direction of motion) manifest themselves as an \emph{oscillatory}, yet \emph{irregular} time dependence, as may be verified from Fig.~\ref{FIG:Diffusion_single} where we plot the MSD of a single system (without averaging).
Averaging over many systems then finally leads to the observed diffusive behavior.
\subsection{Rheological properties \label{SEC:STRESS}}
\begin{figure}
\includegraphics[width=1.0\linewidth]{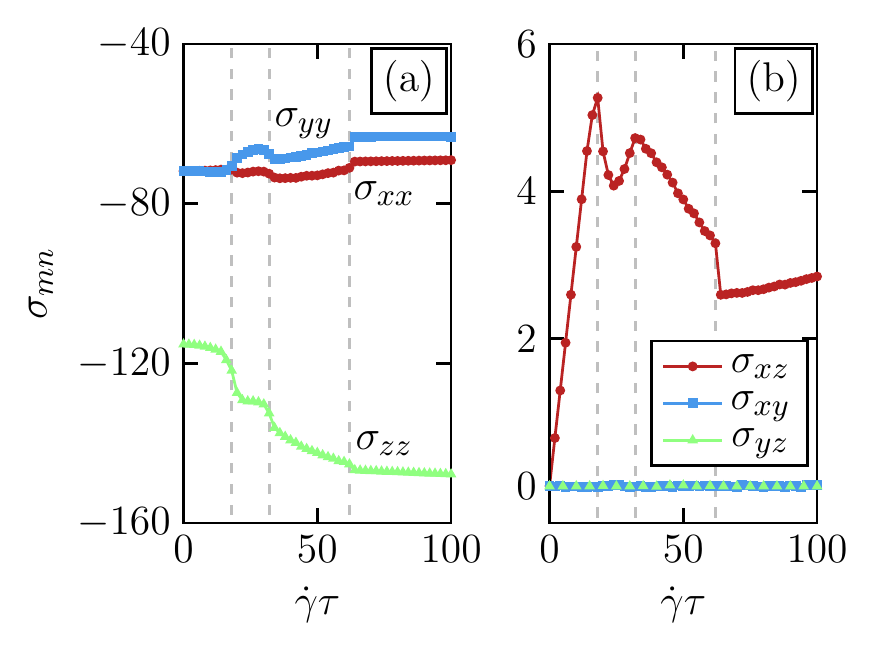}
\caption{(Color online) Components of the stress tensor as a function of the applied shear rate.}
\label{FIG:Stress}
\end{figure}

To complete the picture of the three-layer system we have also investigated
rheological properties as probed by the components $\sigma_{mn}$ (with $n,m \in {x,y,z}$) of the stress tensor $\boldsymbol{\sigma}$. The components were calculated (consistent with \cite{Vezirov2015}) via the virial expression
\begin{align}
\sigma_{mn}&=\left \langle \frac{1}{V} \sum_i \left[ \sum_{j<i} F_{m, ij}n_{ij} + \delta_{nz}\delta_{mz}\, \sigma_w\rb{z_i} \right]\right \rangle,
\label{EQN: STRESS}
\end{align}
where $\delta_{nm}$ is the Kornecker delta, $F_{m,ij}$ is the $m$-component of the two-particles interaction forces, $n_{ij}$ is the $n$-component of $\mb{r}_{ij}$, and $\sigma_w$ is the stress contribution from the confining walls.
The latter is defined as
\begin{align}
  \sigma_w \rb{z_i}= -\frac{\partial u_\text{wall}^{ \rb{\pm} } \rb{z_i} }{\partial z_i } \rb{ z_i \pm L_z/2 }.
\end{align}
The kinetic contributions were neglected.
In fact, in the framework of \emph{overdamped} BD simulations, the only non-vanishing contributions to the kinetic stress are given by the ideal gas contribution, which are $\sigma_{nn}^\text{ideal} = -P = -N \, k_B T / V$ for all diagonal components and zero for all other.
Results for the steady-state stress tensor [i.e., $\lim_{t\to\infty}\sigma_{mn}(t)$] as functions of the dimensionless shear rate are presented in Fig.~\ref{FIG:Stress}.
We consider first the diagonal components (a), the negative values of which correspond to the pressure tensor components of the confined system.
The equilibrium limit ($\dot\gamma\rightarrow 0$) is characterized by $\sigma_{xx}=\sigma_{yy}$, as expected in the fully symmetric quadratic state forming our starting configuration.
The corresponding normal stress $\sigma_{zz}$ is much larger in magnitude, a generic effect in strongly confined fluids.
Upon increasing the shear rate we find that all diagonal components $\sigma_{mm}$ reflect, to some degree, the shear-induced transitions.
As a general trend, the normal stress further increases in magnitude.
We understand this as a consequence of the fact that, with increasing $\dot\gamma$, the distance with the layers progressively increases (see the density profiles in Fig.~\ref{FIG:Density}).
This enables the particles to follow the flow more efficiently.
At the same time, however, it brings more and more particles into the contact zone close to the walls which, consequently, increases the overall repulsion.
A further feature upon increasing $\dot\gamma$ is that, beyond the threshold towards the laned state, $\sigma_{xx}$ and $ \sigma_{yy}$ deviate from one another, indicating a structural asymmetry between flow and vorticity direction.

We now turn to the non-diagonal components [see Fig.~\ref{FIG:Stress}(b)]. The shear stress $\sigma_{xz}$ reveals a strongly non-monotonic behavior with several multivalued regimes where the shear rate corresponding to a certain stress value is not uniquely defined (as already reported earlier by us \cite{Vezirov2015}).
In particular, the ranges of $\dot\gamma$ where the stress decreases imply that the corresponding (laned or shear-molten) state is mechanically unstable and would not be observable in a simulation (or experiment) at constant stress.
The other off-diagonal components ($\sigma_{xy}$, $\sigma_{yz}$) remain zero for all shear rates.

\begin{figure}
\includegraphics[width=1.0\linewidth]{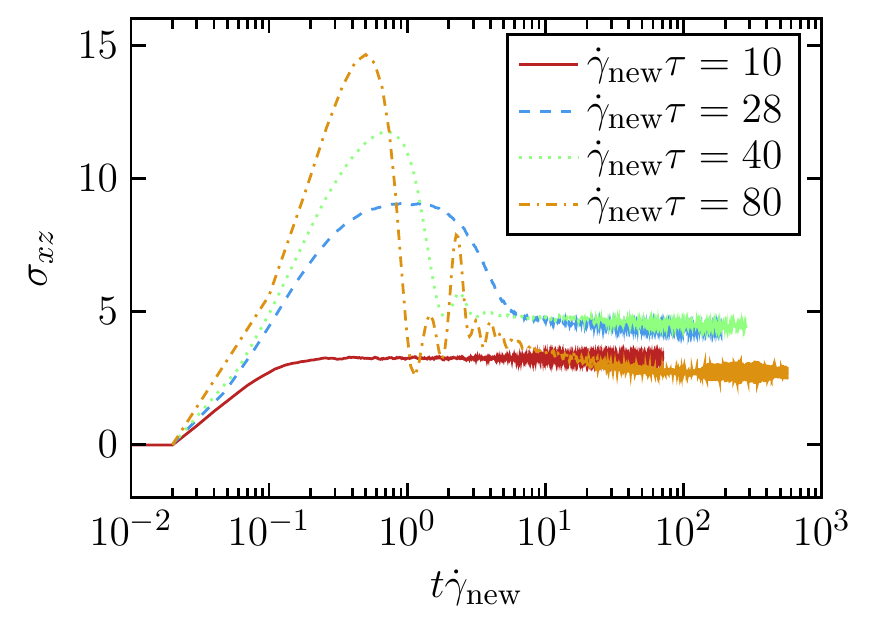}
\caption{ (Color online) Stress-strain relations in the colloidal trilayer for different shear rates $\dot\gamma_\text{new}$, starting from the equilibrium (square)
configuration.
The shear is switched on at $t/\tau = 0.01$. }
\label{FIG:Strain}
\end{figure}

Finally, we consider the relaxation of the (shear) stress after a sudden switch-on of a finite shear rate $\dot\gamma_\text{new}$, starting from the equilibrium (square) configuration at $\dot\gamma=0$.
To this end we plot in Fig.~\ref{FIG:Strain} the quantity $\sigma_{xz}$ as function of the strain $\dot\gamma t$ (with fixed $\dot\gamma=\dot\gamma_\text{new}$).
Similar to earlier investigations made for a bilayer system \cite{Vezirov2015}, the shape of the stress-strain relation strongly depends on the state into which the quench is performed.
In particular, for all shear rates beyond the quadratic regime we observe a pronounced stress "overshoot", i.e., a maximum in the stress-strain relation.
The strain corresponding to this maximum depends on $\dot\gamma_\text{new}$; at the highest value considered it is about ten percent.
This value can be related to the typical size of the nearest-neighbor cages in the quadratic state which are broken by the shear \cite{Frahsa2013}.
We also note that the time in which the stress relaxes towards its final, steady state value is the larger, the closer $\dot\gamma_\text{new}$ is to the "critical" value ($\dot\gamma_c\tau \approx 18$) separating quadratic and laned state.

\section{Conclusions \label{SEC:CONC}}
Using BD simulations we have studied the structural and dynamical behavior of a thin colloidal film confined to a narrow slit-pore under planar shear flow (constant shear rate).
We focused on a dense, strongly confined system whose equilibrium configuration consists of three, well-pronounced layers with square-like crystalline in-plane structure.
Compared to the previously studied case of two layers \cite{Vezirov2013}, the three-layer systems displays a novel dynamical state at shear rates beyond the depinning transition (i.e., the breakdown of the equilibrium structure).
In this "micro-laned" state, the middle layer splits into two sublayers with the particles being organized into lanes with opposite velocities.
Closer inspection shows that the sublayers are "dragged" by the closest (bottom or top) outer layer, which essentially follow the externally applied flow profile.
Overall, the situation corresponds to a plug flow with two domains, whose interface is within the middle layer.
Upon further increase of the shear rate, the system enters a molten state and finally recrystallizes into an ordered state with hexagonal in-plane structure reminiscent of the behavior of the bilayer films \cite{Vezirov2013}.

By monitoring the components of the stress tensor, we find that the micro-laned state is characterized by a negative slope of the shear stress as function of the shear rate.
This suggests that the micro-laned state is, in fact, mechanically unstable and will not be observable in simulations (or experiments) at constant stress.

A similar phenomenon has been reported within an experimental study of strongly confined hard-sphere suspensions \cite{Cohen2004}.
Specifically, the authors of Ref.~\cite{Cohen2004} observe a shear-induced transition from an (initial) configuration consisting of four flat layers to a configuration consisting of three buckled layers, which are characterized by distinct velocities (the layers were distinguished by monitoring their velocity instead of their density profile).
This latter configuration is explained by a "mismatch" of the osmotic pressure of the bath and the pressure (tensor) in the confined system corresponding to flat layers, which need to be balanced.

In the light of these experimental findings, the occurrence of the micro-laned state reported here may be interpreted as a transition from three flat layers to two buckled layers with different mean velocities (see Fig.~\ref{FIG:Velocity}).
From that point of view, our results are qualitatively consistent with the observations of the aforementioned experiments.
A detailed comparison is difficult since the experiments are carried out with the confined system being connected to a bulk suspensions at constant osmotic pressure, as well as at oscillatory shear.
Still, the qualitative agreement may be taken as an indication that the shear-induced laned state is a generic feature of \emph{thin} films with three or more layers.

For future simulation studies, one important direction is to investigate the robustness of the micro-laned state for a larger parameter space. This concerns in particular the slit-pore width and the overall density.
Already in equilibrium, variation of these parameters yields a complex phase diagram \cite{Fortini2006} with different crystalline structures.
The impact of shear on these structures as function of pore width and density has yet to be explored.

Another key question concerns the impact of hydrodynamic interactions on the "micro-laned" state.
The formation of lanes could be significantly enhanced or disturbed by the explicit introduction of solvent dynamics (which was neglected here), due to the small distances between the lanes and rather large velocities.
However, in view of the similarity of the micro-laned state to the buckled state observed in experiments \cite{Cohen2004}, we expect hydrodynamic interactions to affect the time scales, but not the overall behavior of the system.

\begin{acknowledgments}
This work was supported by the Deutsche Forschungsgemeinschaft through SFB 910 (project B2).
\end{acknowledgments}
%

\end{document}